\documentclass[preprint,showpacs,preprintnumbers,amsmath,amssymb]{revtex4}
\usepackage{graphicx}
\usepackage{enumerate}
\usepackage{dcolumn}
\usepackage{bm}
\begin{document}
\title{Calculation of ground- and excited-state energies of confined helium atom}
\author{Arup Banerjee$^{a}$, C. Kamal$^{b}$ and Avijit Chowdhury$^{c}$}
\affiliation{(a) Laser Physics Application Section,
(b) Semiconductor Laser Section, (c) Laser Plasma Lab \\
Centre for Advanced Technology\\
Indore 452013, India}
\begin{abstract}
We calculate the energies of ground and three low lying excited states of confined helium atom centered in an impenetrable spherical box. We perform the calculation by employing variational method with two-parameter variational forms for the correlated two-particle wave function. With just two variational parameters we get quite accurate results for both ground and excited state energies.
\end{abstract}
\maketitle
\section{Introduction}
Recently considerable attention has been focused on the study of spatially confined atoms and molecules \cite{dolmatov,connerade,jaskolski}. The confined atomic and molecular systems show substantially different ground state and response properties as compared to their free counterparts. The main reason for the spatially confined models of atoms and ions to attract tremendous amount of attention is their applicability to several problems of physics and chemistry. For example, atoms trapped in cavities, zeolite channel \cite{tang,frank} or encapsulated in hollow cages of carbon based nano-materials, such as endohedral fullerenes \cite{connerade1,connerade2} and in nanobubbles formed around foreign objects in the environment of liquid helium and under high pressure in the walls of nuclear reactors  \cite{walsh} are all relevant to the confined atom model. The models of confined atomic and molecular system have also found application in the investigation of effect of high pressure on the physical properties of atoms, ions and molecules \cite{michels,sommerfeld}. The study of confined atoms also provides insight into various properties of the quantum nanostructures like quantum dots or artificial atoms \cite{sako1,sako2}. The detail discussion on these applications are available in several review articles \cite{dolmatov,connerade,jaskolski}. 

The first model of confined (compressed) hydrogen atom in an impenetrable spherical cavity was proposed by Michels et al. \cite{michels} to simulate the effect of pressure on hydrogen atom and this model was employed to  study the variation of static dipole polarizability with the effective pressure acting on the surface. In this model the boundary condition that the wave function vanishes at $r = r_{c}$ ( where $r_{c}$ is the radius of impenetrable spherical box) is imposed on the solution of the Schrodinger equation. Later various physical properties of confined hydrogen atom, such as the modification of their atomic orbitals, energy levels, the filling of electronic shells, linear and nonlinear polarizabilities have been reported in the literature ( see Ref. \cite{dolmatov} and references theirin). Besides hydrogen atom, effect of confinement by an impenetrable as well as non-impenetrable spherical box, on many-electron atoms have also been considered \cite{tenseldam,gimarc,ludena1,ludena2,martin,joslin,aquino,patil}. Helium atom being the simplest many-electron system, the confined version of this atom provides a lucid way to study the effect of confinement on the electron correlation which arises due to the coulomb interaction between the two electrons. Most of the studies on the confined helium atom are devoted to the calculation of the ground state energies and some averages and their evolution with the size of the spherical box. In majority of these studies Raleigh-Ritz variational method was employed with modified Hylleraas-type wave function (Hylleraas-type wave functions multiplied with appropriate cut-off factor) fulfilling the confinement boundary condition mentioned above \cite{tenseldam,gimarc,aquino}. Besides variational method, self consistent Hartree-Fock \cite{ludena1}, configuration interaction \cite{ludena2} and a quantum Monte Carlo (QMC) \cite{joslin} methods  have also been used to study the properties of helium atom and several isoelectronic ions confined in an impenetrable spherical box. 

In this paper we report calculations of energies of the ground state and some low lying excited states of helium atom confined at the center of an impenetrable spherical box and study the variation of the energies with the size of the sphere. We note here that in comparison to the calculation of ground state energy very few studies on the evolution of excited state energies with the size exist in the literature \cite{varshni,patil}. Therefore, main emphasis of the present paper is on the effect of confinement on some of the low lying excited states of compressed helium atom.   The low lying excited states considered in this paper are $^{3}S (1s2s)$, $^{1}P(1s2p)$ and $^{3}P(1s2p)$. Recently, Patil and Varshni \cite{patil} have calculated the the energies of the above mentioned excited states by replacing the electron-electron interaction by an effective screening of the nuclear charge. The screening factor is then determined by using an interpolation between the expressions for large and small values of the confining radius. In the present paper calculations are performed by employing more accurate variational method with two-parameter correlated wave functions for both ground and the excited states. The correlated wave functions explicitly take the effect of electron-electron interaction into account and consequently expected to yield accurate results. The variational forms for the wave functions we use in this paper are generalization of the correlated wave functions proposed by Le Sech and co-workers \cite{sech1,sechcpl,sech2,sech3} for free two-electron atomic and molecular systems. The generalized wave functions for calculations of confined helium atom are constructed by multiplying the Le Sech type wave functions with appropriate cut-off factors so that the confinement boundary condition is satisfied. In addition to the confinement boundary condition these wave functions also fulfill both electron-nucleus and electron-electron cusp conditions. The cusp conditions arise due to the Coulomb interaction between the charged particles and the true wave functions of many electron systems must satisfy these conditions  \cite{myers}.  At this point it is important to note that all previous variational calculations involved more than two variational parameters. We demonstrate that the two-parameter calculations performed in this paper yield quite accurate results and match well with the results of some accurate calculations already exist in the literatures. 
   
The remaining paper is organized in the following manner. In section II we describe the theoretical methods employed in this paper. The section III is devoted to the discussion of the results. The paper is concluded in section IV.
\section{Method of Calculation}
The non-relativistic Schr$\ddot{o}$dinger equation for confined two-electron helium-like systems with nuclear charge Z can be written as ( in atomic units)
\begin{equation}
\left [ -\frac{1}{2}{\vec{\nabla}}_{1}^{2} - \frac{1}{2}{\vec{\nabla}}_{2}^{2} + v_{N}({\bf r}_{1},{\bf r}_{2}) + v_{C}({\bf r}_{1},{\bf r}_{2}) + v_{conf}({\bf r}_{1},{\bf r}_{2})\right ]\psi({\bf r}_{1},{\bf r}_{2}) = E\psi({\bf r}_{1},{\bf r}_{2})
\label{schrodinger}
\end{equation}
where $v_{N}$ is the nuclear potential
\begin{equation}
v_{N}({\bf r}_{1},{\bf r}_{2}) = -\frac{Z}{r_{1}} - \frac{Z}{r_{2}},
\label{nucpot}
\end{equation}
$v_{C}$ represents coulomb repulsion between the electrons
\begin{equation}
v_{C}({\bf r}_{1},{\bf r}_{2}) = \frac{1}{r_{12}},
\label{coulpot}
\end{equation}
and the confining potential $v_{conf}$ due to an impenetrable spherical box of radius $r_{c}$ is given by
\begin{equation}
v_{conf}({\bf r}_{1},{\bf r}_{2}) =
\left\{\begin{array}{cl}
0 & \mbox{$r_{1}, r_{2} < r_{c}$} \\
\infty & \mbox{$r_{1}, r_{2}\geq r_{c}$}
\end{array}\right.
\label{cofpot}
\end{equation}
To solve the above Schr$\ddot{o}$dinger equation for the both ground and excited states we employ Raleigh-Ritz variational approach by finding the stationary solutions of the following energy functional
\begin{equation}
E[\psi ] = \frac{\int\psi^{*}({\bf r}_{1},{\bf r}_{2})\left [-\frac{1}{2}{\vec{\nabla}}_{1}^{2} - \frac{1}{2}{\vec{\nabla}}_{2}^{2} + v_{N} + v_{C} + v_{conf}\right ]\psi({\bf r}_{1},{\bf r}_{2})d{\bf r}_{1}d{\bf r}_{2}}{\int\psi^{*}({\bf r}_{1},{\bf r}_{2})\psi({\bf r}_{1},{\bf r}_{2})d{\bf r}_{1}d{\bf r}_{2}}
\label{enrgyfunc1}
\end{equation}
In order to perform variational calculation we need to make a judicious choice for the  ansatz of two-particle wave function. To this end we generalize the variational form for wave function  $\psi({\bf r}_{1},{\bf r}_{2})$ proposed by Le Sech and co-workers \cite{sech1,sechcpl,sech2,sech3}. The generalized variational form of the wave function, complying with the boundary condition imposed by the confining potential, employed in this paper is given by.
\begin{equation}
\psi({\bf r}_{1},{\bf r}_{2}) = \phi({\bf r}_{1},{\bf r}_{2})\Omega({\bf r}_{1},{\bf r}_{2})\left (1 - \frac{r_{1}^{2}}{r_{c}^{2}}\right )\left (1 - \frac{r_{2}^{2}}{r_{c}^{2}}\right ) 
\label{varwav1}
\end{equation} 
This form of the wave function is inspired by the concept of semi separability, introduced by Pluvinage \cite{pluvinage}. Following Ref. \cite{sech1,sechcpl,sech2,sech3} $\phi({\bf r}_{1},{\bf r}_{2})$  is chosen in such a way that it only includes the wave function of the free electrons in the field of nucleus and $\Omega({\bf r}_{1},{\bf r}_{2})$ contains the dependence of inter-electronic distance $r_{12}$ needed to represent the correlation arising due to coulomb interaction between the electrons.  Moreover, the presence of cut-off factor $\left (1 - \frac{r_{1}^{2}}{r_{c}^{2}}\right )\left (1 - \frac{r_{2}^{2}}{r_{c}^{2}}\right )$ ensures the satisfaction of boundary condition imposed by the confining potential on the wave function.  In addition to the satisfaction of confining boundary condition the above form of the cut-off factor also guarantees the fulfillment of the electron-nucleus cusp condition by the correlated wave function. The satisfaction of cusp condition by the variational wave function leads to improvement in the rate of convergence of Rayleigh-Ritz variational calculations and also yields more accurate energies \cite{myers}. We note here that the calculations performed with linear form of the cut-off factor $\left (1 - \frac{r_{1}}{r_{c}}\right )\left (1 - \frac{r_{2}}{r_{c}}\right )$, leading to trial wave functions that do not satisfy electron-nucleus cusp condition, give less accurate results for the ground as well as the excited state energies, specially in the strong confinement regime. Consequently, in this paper we employ the quadratic form of the cut-off factor as given by Eq. (\ref{varwav1}).  For detail discussion on the various asymptotic and cusp properties satisfied by the Le Sech form of the wave function we refer the reader to Ref. \cite{sech3}. With this choice for the variational form of the correlated wave function, the energy functional (\ref{enrgyfunc1}) reduces to a single multidimensional quadrature
\begin{equation}
E[\phi{\tilde\Omega}] = E_{0} + \int\phi^{2}\left [\frac{{\vec{\nabla} }_{1}{\tilde\Omega}\cdot{\vec{\nabla}}_{1}{\tilde\Omega} + {\vec{ \nabla}}_{2}{\tilde\Omega}\cdot{\vec{\nabla}}_{2}{\tilde\Omega}}{2} + \frac{{\tilde\Omega}^{2}}{r_{12}}\right ]d{\bf r}_{1}d{\bf r}_{2}
\label{energyfunc2}
\end{equation}
where ${\tilde\Omega} = \Omega\left (1 - \frac{r_{1}^{2}}{r_{c}^{2}}\right )\left (1 - \frac{r_{2}^{2}}{r_{c}^{2}}\right )$ and $E_{0}$ denotes the energy of two-electron systems moving in the nuclear potential only.  
The forms of the functions $\phi({\bf r}_{1},{\bf r}_{2})$ and $\Omega({\bf r}_{1},{\bf r}_{2})$ for the ground and some excited states employed in this paper for calculating corresponding energies are presented in the next section along with the results they yield for confined helium atom.
\section{Results and Discussion} 
In this section we present the results of our calculations and compare them with the results already available in the literature. First we discuss the results for the ground state energy and its variation with the size of the spherical box followed by the results for some of the low lying excited states of confined helium atom.
\subsection{Ground state}
The form of the function $\phi({\bf r}_{1},{\bf r}_{2})$ for the ground state $^{1}S(1s^{2})$ of helium atom is written as
\begin{equation}
\phi({\bf r}_{1},{\bf r}_{2}) = A\left (1s({\bf r}_{1})1s({\bf r}_{2})\right )\left [\alpha(1)\beta(2) - \alpha(2)\beta(1)\right ]
\end{equation}
where $A$ is the normalization constant, $1s({\bf r})$ represents the ground state hydrogenic
orbital and $\alpha(i)$ and $\beta(j)$ denote the spin-up and spin-down functions respectively. To perform the calculations we next choose the form of function $\Omega(r_{1},r_{2},r_{12})$, which describes the correlation between the electrons arising due to the coulomb interaction as
\begin{equation}
\Omega(r_{1},r_{2},r_{12}) = \cosh\lambda r_{1}\cosh\lambda r_{2}\left (1 + \frac{1}{2}r_{12}e^{-ar_{12}}\right )
\label{hirsh}
\end{equation}
where $\lambda$ and $a$ are the two variational parameters which are determined by minimization of the energy functional (\ref{energyfunc2}). The above form of $\Omega(r_{1},r_{2},r_{12})$ consists of two parts, namely, the screening part represented by the cosh hyperbolic functions and purely $r_{12}$ dependent correlated part.  The $r_{12}$ part of the the function was proposed by Hischfelder \cite{hirschfelder}  and its applicability has been demonstrated for several many-electron systems up to beryllium atom \cite{sech1,sechcpl,sech2,sech3,sech4}. This function provides right description of electron-electron cusp condition. In accordance with Ref. \cite{sech3} we choose to represent the screening part of $\Omega(r_{1},r_{2},r_{12})$  by cosh hyperbolic functions which fulfill the cusp condition at the nucleus and also right behaviour at large electron-nucleus distances. However, in contrast to the Ref. \cite{sech3} we represent the screening part by product of cosh hyperbolic functions at $r_{1}$ and $r_{2}$ instead of the sum of two functions. Although the product form of the screening part considered in this paper overestimates the screening, nonetheless we find that this form leads reasonably accurate result for the ground state energy of an uncompressed or a free helium atom.  The difference between our result for the ground state of uncompressed helium atom and the one obtained with sum of two cosh hyperbolic functions \cite{sech3} is of the order of $0.002$ a.u.. Having assessed the accuracy of our ansatz for the ground state energy of free helium atom we now proceed with the calculation of energies for confined helium atom as a function of size of the confining spherical box.

In Table I we present the results of our calculations for energies of $^{1}S(1s^{2})$ state of helium atom as a function of $r_{c}$ along with the corresponding results available in the literature. In order to check the accuracy of our results we make a comparison with the results of some very accurate calculations based on correlated wave function  \cite{gimarc,aquino} and QMC \cite{joslin} approaches. It can be clearly seen from Table I that our results are slightly lower than the corresponding numbers of Ref. \cite{gimarc} especially for small values of $r_{c}$. It is important to note here that in Ref. \cite{gimarc} ground state energies were obtained by employing three-parameter variational calculation with different form for the function of $r_{12}$. On the other hand our results are slightly higher than the results obtained both with the thirteen-parameter (ten linear and three nonlinear) Hylleraas wave function variational calculation \cite{aquino} and the most accurate QMC based method.  In the strong confinement regime (that is for small values of $r_{c}$) the maximum difference between our results and those of Ref. \cite{joslin,aquino} is of the order of $0.006$ a.u.. The results presented in Table I clearly demonstrate that the two-parameter ansatz used in this paper to calculate the ground state energies of confined helium atom gives quite accurate results for wide range of size of the confining spherical box.
\subsection{Excited states}
Now we apply the variational approach discussed above to calculate the energies of some low lying singly excited states and study their variations with the size of the spherical box. Following, Ref. \cite{sech1,sech2} the $\phi({\bf r}_{1},{\bf r}_{2})$ part of the correlated wave function for the excited state $1snl$ (where $n$ is the principal quantum and $l$ is the orbital angular momentum quantum number of the state to which one electron is excited) is chosen as
\begin{equation}
\phi({\bf r}_{1},{\bf r}_{2}) = A\left [1s({\bf r}_{1})nl({\bf r}_{2}) \pm 1s({\bf r}_{2})nl({\bf r}_{1})\right ]\chi_{s}(1,2)
\label{phiexcit}
\end{equation}
where $nl({\bf r})$ is hydrogenic orbital with quantum number $n$ and $l$ and $\chi_{s}(1,2)$ represents the spinor part of the wave function. The spinor part can be easily constructed by using spin-up $\alpha(i)$ and spin-down $\beta(j)$ functions such that the total wave function should be antisymmetric with respect to the interchange of spatial and spin co-ordinates of two electrons. For calculation of energies of excited states the screening cum $r_{12}$ dependent part of the wave function is chosen as 
\begin{equation}
\Omega(r_{1},r_{2},r_{12}) = \left (\cosh\lambda r_{1} + \cosh\lambda r_{2}\right )\left (1 + \frac{1}{2}r_{12}e^{-ar_{12}}\right )
\label{hirshex}
\end{equation} 
The screening part in the above equation (Eq. (\ref{hirshex})) is motivated by the work of Ref. \cite{sechcpl}. Note that the form of screening part of the wave function for excited state calculation is different from that of ground state calculation. We find that product form leads to less accurate results for the excited state of  uncompressed or free helium atom. Consequently, for excited state calculation we employ the above form (Eq. (\ref{hirshex})) which has already been shown to yield accurate results for the excited states of two-electron systems \cite{sechcpl}. By using $\phi({\bf r}_{1},{\bf r}_{2})$ given by Eq. (\ref{phiexcit}) we calculate energies of some low lying excited states for which the values of quantum number $l$ or different spin state automatically ensures the orthogonality. These states are $^{3}S (1s2s)$, $^{1}P(1s2p)$ and $^{3}P(1s2p)$. Before presenting the results for the confined helium atom we note that with Eqs. (\ref{phiexcit}) and (\ref{hirshex}) we get quite accurate results for the excited states of uncompressed helium atom. For excited state $^{3}S (1s2s)$ of uncompressed helium atom we get $E = -2.1743$ a.u. in comparison to the most accurate value of $E = -2.1752$ a.u.  \cite{frolov1} (here we quote up to fourth-decimal place only). On the other hand, our calculation for $^{1}P(1s2p)$ and $^{3}P(1s2p)$ states of uncompressed helium atom give $E = -2.1227$ a.u. and $E = -2.1281$ a.u. respectively, whereas the corresponding values of energies from accurate calculations are $E = -2.1238$ a.u. and $E = -2.1332$ a.u.  \cite{frolov2} (here we quote up to fourth-decimal place only) respectively. The results for energies of three excited states mentioned above and their variations with respect to $r_{c}$ are presented in Table II along with the corresponding results from Ref. \cite{patil} in parenthesis. The comparison of two results clearly shows that our numbers for $^{3}S(1s2s)$ states are lower than the corresponding number of Ref. \cite{patil} for almost whole range of $r_{c}$ considered in this paper except for $r_{c} = 1.0$ a.u.. Moreover, for $^{3}S(1s2s)$ state two results differ significantly in the range $r_{c} = 2.0 - 9.0 a.u.$. In contrast to $^{3}S(1s2s)$ state case, our results for  $^{1}P(1s2p)$ state are slightly lower than the corresponding numbers of Ref. \cite{patil} for all values of $r_{c}$ excepting $r_{c} = 1.0$ a.u.. Finally, we note from Table II that unlike excited states $^{3}S(1s2s)$ and $^{1}P(1s2p)$ our results for the excited state $^{3}P(1s2p)$ are lower than the corresponding results of Ref. \cite{patil} only in the range $r_{c} = 2.0 - 9.0$ a.u.. Beyond $r_{c} = 9.0$ a.u. results obtained by us are little higher than those of Ref. \cite{patil}. The lower values of energies for excited states obtained by our variational calculation particularly in the strong confinement regime suggests that our results may be more accurate than those of Ref. \cite{patil}. 
\section{Conclusion}
In this paper we have calculated the energies for ground and some excited states of confined helium atom by employing Raleigh-Ritz variational method. To perform the variational calculation we have used two-parameter variational forms for two-electron correlated wave function taking into account the boundary condition imposed by the confinement. The results obtained by us for the ground state is quite accurate within the strong confinement region  
and match well with the corresponding results of other accurate calculations. For excites states not many results are available in the literature. We have made comparison of our results for the excited states with the numbers from an interpolation based approach published recently. The comparison of results shows that our variation method with two-parameter wave function is capable of giving quite accurate numbers for the excited states of confined helium atom for a wide range of values of confinement radius. We feel that our results for the excited states will be an important contribution to the spectroscopic properties of confined helium atom and accuracy of these results can be further tested by more sophisticated calculations. The study of confined two-electron systems can be extended in several directions. The calculations of response properties like linear and nonlinear polarizabilities will provide information about the interaction of confined system with external electromagnetic field. Currently we are studying the effect of confinement on the linear and nonlinear polarizabilities of confined helium atom. 
\section{acknowledgement} We wish to thank Prof. Manoj Harbola and Dr. Selva Nair for useful discussions.

\newpage
\begin{table}
\caption{Energies for the ground state $^{1}S (1s^{2})$ of confined helium atom as a function of spherical box radius.  All numbers are in atomic units.}
\tabcolsep=0.3in
\begin{center}
\begin{tabular}{|c|c|c|c|c|}
\hline
$r_{c}$ & Present & Ref. \cite{gimarc} & Ref. \cite{aquino} & Ref. \cite{joslin} \\
\hline
0.6 & 13.3343 & - & 13.3183 & - \\
0.7 & 7.9320 & 7.9491 & 7.9255 & -  \\
0.8 & 4.6157 & 4.6225 & 4.6106 & - \\ 
0.9 & 2.4670 & 2.4691 & 2.4633 & -  \\ 
1.0 & 1.0183 & 1.0186 & 1.0159 & 1.0142 \\ 
1.1    & 0.0106 & 0.0106 & 0.0091 & -\\
1.2 & -0.7079 & -0.7075 & -0.7087 & - \\
1.3    & -1.2304 & -1.2295 &-1.2309 & -\\
1.4 & -1.6167 & -1.6151 & -1.6172 & - \\ 
1.5    & -1.9061 & -1.9040 & -1.9067 & 1.9081 \\
1.6 & -2.1253 & -2.1229 & -2.1263 & - \\   
1.7    & -2.2928 & -2.2903 & -2.2944 & - \\
1.8 & -2.4219 & -2.4193 & -2.4242 & -  \\
1.9    & -2.5219 & -2.5195 & -2.5249 & - \\
2.0 & -2.5998 & -2.5977 & -2.6036  &  -2.6051 \\
2.2     & -2.7088 & -2.7074 & -2.7141 & - \\ 
2.4 & -2.7765 & -2.7760 & -2.7831 & -  \\
2.6     & -2.8191 & -2.8194 & -2.8266 & - \\
2.8 & -2.8462 & -2.8472 & -2.8542  & - \\
3.0     & -2.8636 & -2.8652 & -2.8718 & -2.8727\\
3.5 & -2.8851 & - & -2.8928 & -2.8935\\
4.0    & -2.8931 & -2.8956 & -2.8997 & -2.9003\\
4.5 & -2.8963 & - & -2.9020  & - \\   
5.0     & 2.8978 & -2.9004 & -2.9028& - 2.9032\\
5.5 & -2.8985 & - & -2.9031  & - \\   
6.0 & -2.8990 & - & -2.9033 & -2.9035\\
$\infty$ & -2.8999 & -2.9024 & -2.9035 & -2.9037\\
\hline
\end{tabular}
\end{center}
\end{table} 

\begin{table}
\caption{Energies for three low lying excited states of confined helium atom as a function of spherical box radius. Numbers in parenthesis are results of Ref. \cite{patil}. All numbers are in atomic units.}
\tabcolsep=0.5in
\begin{center}
\begin{tabular}{|c|c|c|c|}
\hline
$r_{c}$ & $^{3}S (1s2s)$ & $^{1}P(1s2p)$ & $^{3}P(1s2p)$  \\
\hline
1.0 & 15.5451 & 8.0312 & 7.5265  \\
     & (15.050) & (7.751) & (7.680) \\
2.0 & 0.5862 & -0.3414 & -0.4907  \\
     & (0.9809) & (-0.3334) & (-0.3692) \\
3.0 & -1.3679 & -1.5217 & -1.5758  \\ 
    & (-1.1193) & (-1.5069) & (-1.5312) \\ 
4.0 & -1.8734 & -1.8598 & -1.8907  \\ 
     & (-17277) & (-1.8499) & (-1.8686) \\
5.0 & -2.0473 & -1.9928 & -2.0126  \\ 
    & (-1.9615) & (-1.9857) & (-2.0012) \\
6.0 & -2.1171 & -2.0547 & -2.0684  \\
    & (-2.0658) & (-2.0489) & (-2.0624) \\
7.0 & -2.1477& -2.0861 & -2.0964  \\ 
    & (-2.1166) & (-2.0812) & (-2.0934) \\
8.0 & -2.1617 & -2.1029 & -2.1111  \\   
    & (-2.1429) & (-2.0987) & (-2.1101) \\
9.0 & -2.1683 & -2.1121 & -2.1191 \\
    & (-2.1570) & (-2.1087) & (-2.1195) \\
10.0 & -2.1714 & -2.1172 & -2.1234  \\
     & (-2.1647) & (-2.1146) & (-2.1249) \\ 
11.0 & -2.1729 & -2.1199 & -2.1257  \\
     & (-2.1691) & (-2.1181) & (-2.1281) \\
12.0 & -2.1736 & -2.1215 & -2.1269  \\
     & (-2.1716) & (-2.1202) & (-2.1300) \\
13.0 & -2.1740 & -2.1223 & -2.1276  \\
     & (-2.1731) & (-2.1215) & (-2.1312) \\
14.0 & -2.1742 & -2.1227 & -2.1280 \\   
     & (-2.1739) & (-2.1223) & (-2.1320) \\ 
\hline
\end{tabular}
\end{center}
\end{table}
\end{document}